\documentclass{webofc}
\usepackage[varg]{txfonts}
\usepackage{latexsym,graphicx,epsfig,psfrag,here}
\usepackage{epstopdf}
\usepackage{amssymb,amsmath,amsxtra,amsfonts}
\usepackage{bm}

\usepackage{longtable}
\usepackage{multirow}
\usepackage{booktabs}
\usepackage{array}
\usepackage{wrapfig}
\usepackage{color}
\usepackage{titlesec}

\newcommand{\bea}{\begin{eqnarray}}
\newcommand{\ena}{\end{eqnarray}}
\newcommand{\be}{\begin{equation}}
\newcommand{\en}{\end{equation}}
\newcommand{\nn}{\nonumber\\}
\newcommand{\ed}{\end{document}}

\newcommand{\bl}{\bigl}
\newcommand{\br}{\bigr}
\newcommand{\la}{\langle}
\newcommand{\ra}{\rangle}

\begin{document}

\title{Semileptonic decays of $ B_c$ mesons into charmonium states}

\author{\firstname{Aidos} \lastname{Issadykov}\inst{1,2}\fnsep\thanks{\email{issadykov@jinr.ru}} \and
        \firstname{Mikhail A.} \lastname{Ivanov}\inst{1} \and
        \firstname{Guliya} \lastname{Nurbakova}\inst{3}
}

\institute{Joint Institute for Nuclear Research, Dubna, Russia
\and
           The Institute of Nuclear Physics,Ministry of Energy of the Republic of Kazakhstan, Almaty,Kazakhstan 
\and
          Al-Farabi Kazakh National University, Almaty, Kazakhstan
          }

\abstract{%
In this work we study the semileptonic decays of $B_c$ meson. We evaluated $B_{c}\rightarrow D(D^{\ast})$, $B_{c}\rightarrow D_s(D_s^{\ast})$ and $B_{c}\rightarrow \eta_{c}(J/\psi)$ transitions form factors in the full kinematical
region within the covariant quark model. The calculated form factors are used to evaluate the semileptonic decays of $B_c$ meson and it was defined ratios ($R_{\eta_{c}}$, $R_{J/\psi}$, $R_{ D}$ ,$R_{ D^{\ast}}$) of
the branching ratios, which will be hopefully tested on LHC experiments.We
compare the obtained results with the results from other theoretical approaches.
}

\maketitle

\section{Model}
\label{sec:model}

The covariant quark model was developed by G.V.Efimov and M.A.Ivanov
\cite{Efimov:1988yd,Faessler:2002ut,Branz:2009cd}.

The effective Lagrangian describing
the transition of a meson $M(q_1\bar q_2)$ to its constituent
quarks $q_1$ and $\bar q_2 $ in model looks like  
\bea
{\mathcal L}_{\rm int}(x) &=& g_M M(x)\cdot J_M(x) + {\rm h.c.},
\nn
J_M(x) &=& \int\!\! dx_1 \!\!\int\!\!
dx_2 F_M (x,x_1,x_2)\bar q_2(x_2)\Gamma_M q_1(x_1) ,
\ena
with $\Gamma_M$ a Dirac matrix which projects onto the spin quantum 
number of the meson field $M(x)$. The vertex function $F_M$  characterizes 
the finite size of the meson. Translational invariance requires the 
function $F_M$ to fulfill the identity $F_M(x+a,x_1+a,x_2+a)=F_M(x,x_1,x_2)$ for
any four-vector $a$. A specific form for the  vertex function is adopted
\be
F_M(x,x_1,x_2)=\delta(x - w_1 x_1 - w_2 x_2) \Phi_M((x_1-x_2)^2),
\label{eq:vertex}
\en
where $\Phi_M$ is the correlation function of the two constituent quarks
with masses $m_{q_1}$, $m_{q_2}$ and the mass ratios
$w_i = m_{q_i}/(m_{q_1}+m_{q_2})$.

A simple Gaussian form of the vertex function $\bar \Phi_M(-\,k^2)$ is selected
\be
\bar \Phi_M(-\,k^2) 
= \exp\left(k^2/\Lambda_M^2\right)
\label{eq:Gauss}
\en
with the parameter $\Lambda_M$ linked to the size of the meson. The minus sign 
in the argument is chosen to indicate that we are working in the Minkowski 
space. Since $k^2$ turns into $-\,k_E^2$ in the Euclidean space, 
the form (\ref{eq:Gauss}) has the appropriate fall-off behavior in 
the Euclidean region. Any choice for  $\Phi_M$ is appropriate
as long as it falls off sufficiently fast in the ultraviolet region of
the Euclidean space to render the corresponding Feynman diagrams ultraviolet 
finite. We choose a Gaussian form for calculational convenience.

The fermion propagators for the quarks are given by
\be
S_i(k)=\frac{1}{m_{q_i}-\not\! k}
\label{eq:prop}
\en
with an effective constituent quark mass $m_{q_i}$. 

The so-called {\it compositeness condition} 
\cite{ Weinberg:1962hj, Salam:1962ap} is used 
to determine the value of the coupling constants $g_M$. 
It means that the renormalization constant $Z_M$ of the elementary meson 
field $M(x)$ is to be set to zero, i.e.,
\be
Z_M \, = \, 1 - \, \frac{3g^2_M}{4\pi^2} \,\bar\Pi'_M(m^2_M) \, = \, 0,
\label{eq:Z=0}
\en
where $\bar\Pi^\prime_M$ is the derivative of the meson mass operator.
Its physical meaning in Eq.~(\ref{eq:Z=0}) becomes clear when interpreted as 
the matrix element between the physical and the corresponding bare state: 
$Z_M=0$ implies that the physical state does not contain the bare state and 
is appropriately described as a bound state. The interaction makes 
the physical particle dressed, i.e. its mass and wave function have to be 
renormalized. The condition $Z_M=0$ also effectively excludes the constituent 
degrees of freedom from the space of physical states. It thereby guarantees 
the absence of double counting for the physical observable under consideration,
the constituents exist only in  virtual states. The tree-level diagram together
with the diagrams containing self-energy insertions into the external legs 
(i.e. the tree-level diagram times $Z_M -1$) give
a common factor $Z_M$  which is equal to zero.

The mass functions for the pseudoscalar meson (spin $S=0$)
and vector meson (spin $S=1$) are defined as
\bea
\Pi_{P}(x-y) &=& +\,i\,\la T\bl\{J_P(x)J_P(y) \br\} \ra_0 ,
\label{eq:S=0}\\[2ex]
\Pi^{\mu\nu}_{V}(x-y) &=& -\,i\,\la T\bl\{J^\mu_V(x)J^\nu_V(y) \br\} \ra_0 .
\label{eq:S=1}
\ena

Herein we use the updated values of the model parameters from
\cite{Dubnicka:2016nyy} which are shown in Eq.~(\ref{eq:fit},\ref{eq:fitsize}).
\be
\def\arraystretch{1.5}
\begin{array}{ccccc|c}
     m_{u/d}        &      m_s        &      m_c       &     m_b & \lambda  
\\\hline
 \ \ 0.241\ \   &  \ \ 0.428\ \   &  \ \ 1.67\ \   &  \ \ 5.05\ \   & 
\ \ 0.181\ \      & \ {\rm GeV} 
\end{array}
\label{eq:fit}
\en

\begin{equation}
\def\arraystretch{2}
\begin{array}{cccccccccc|c}
  \Lambda_{B_c} & \Lambda_{\eta_c}			& \Lambda_{J/\psi}	& \Lambda_D        	&  \Lambda_{D^\ast} 
 &\Lambda_{D_s} &  \Lambda_{D^{\ast}_{s}} 	& \Lambda_{B} 		& \Lambda_{B^\ast} 	&  \Lambda_{B_s}& \\
\hline
  \ \ 2.73 \ \  & \ \ 3.97 \ \ 				& \ \ 1.74 \ \ 		&  \ \ 1.6\ \      	& \ \ 1.53 \ \ 
 &\ \ 1.75 \ \  & \ \ 1.56 \ \ 				& \ \ 1.96 \ \ 		& \ \ 1.8\ \      	& \ \ 2.05 \ \ &\  {\rm GeV} \\
\end{array}
\label{eq:fitsize}
\end{equation}

\section{Semileptonic decays}
\label{sec:semdec}
We give
the necessary definitions of the leptonic decay constants,
invariant form factors and helicity amplitudes.
The leptonic decay constants are defined by
\begin{eqnarray}
&&
M(H_{12}\to \bar l \nu)=
\frac{G_F}{\sqrt{2}}\,V_{q_1q_2}\,{\mathcal M}_H^\mu(p)\,
\bar u_l(k_l)\,O^\mu\,u_\nu(k_\nu),
\nonumber\\
&&\nonumber\\
&&
{\mathcal M}_H^\mu(p) = -\,3\,g_{12}\,
\int\!\frac{d^4k}{(2\,\pi)^4\,i}\,
\widetilde\Phi_{12}\left(-k^2\right)\,
{\rm tr}\left[\Gamma_H\,\widetilde S_2(k-c^2_{12}\, p)\,
O^\mu\,\widetilde S_1(k+c^1_{12}\, p)\,\right]\,,
\nonumber\\
&&
\Gamma_P=i\,\gamma^5,\,\,\, \Gamma_V=\varepsilon_V\cdot \gamma,
\nonumber\\
&&\nonumber\\
&&
{\mathcal M}_P^\mu(p)= -i f_P p^\mu,
\hspace{1cm}
{\mathcal M}_V^\mu(p)= f_V m_V \varepsilon_V^\mu.
\label{lept}
\end{eqnarray}

The semileptonic  decays of the $B_c$-meson may be
induced by a b-quark transition.
\begin{eqnarray}
M(H_{13}\to H_{23}+\bar l\nu) & = &
\frac{G_F}{\sqrt{2}}\,V_{q_1q_2}\,{\mathcal M^\mu_{12}}(p_1,p_2)\,
\bar u_l(k_l)\,O^\mu\,u_\nu(k_\nu),
\nonumber\\
&&\nonumber\\
{\mathcal M^\mu_{12}} & = &
-\,3\,g_{13}\,g_{23}\,
\int\!\frac{d^4k}{(2\,\pi)^4\,i}\,
\widetilde\Phi_{13}\left(-(k+c^3_{13}\,p_1)^2\right)\,
\widetilde\Phi_{23}\left(-(k+c^3_{23}\,p_2)^2\right)\,
\nonumber\\
&&
\times{\rm tr}\left[\,i\,\gamma^5\,\widetilde S_3(k)\,
\Gamma_{32}\,\widetilde S_2(k+p_2)\,O^\mu\,\widetilde S_1(k+p_1)\,\right],
\nonumber\\
\nonumber\\
&&
\times{\rm tr}\left[\,i\,\gamma^5\,\widetilde S_3(k-p_1)\,
O^\mu\,\widetilde S_2(k-p_2)\Gamma_{21}\,\widetilde S_1(k)\,\right],
\end{eqnarray}
where $q_1\equiv b$ and
$q_3\equiv c$ whereas $q_2$ denotes either of $c,u,d,s$.

The invariant form factors for the semileptonic $B_c$-decay
into the hadron with spin $S=0,1$ are defined by

\begin{eqnarray}
{\mathcal M^\mu_{\,S=0}} &=&
P^\mu\,F_+(q^2)+q^\mu\,F_-(q^2),\label{ff0}\\
&&\nonumber\\
{\mathcal M^\mu_{\,S=1}}&=&
\frac{1}{m_1+m_2}\,\epsilon^\dagger_\nu\,
\left\{\,
-\,g^{\mu\nu}\,Pq\,A_0(q^2)+P^\mu\,P^\nu\,A_+(q^2)
+q^\mu\,P^\nu\,A_-(q^2)
+i\,\varepsilon^{\mu\nu\alpha\beta}\,P_\alpha\,q_\beta\,V(q^2)\right\},
\label{ff1}\\
&&\nonumber\\
P &=&p_1+p_2, \qquad q=p_1-p_2.
\nonumber
\end{eqnarray}

It is convenient to express all physical observables
through the helicity form factors $H_m$.
The helicity form factors $H_m$ can be expressed in terms of
the invariant form factors in the following way \cite{Ivanov:2000aj}:

\vspace{0.5cm}
\noindent
(a) Spin $S=0$:

\begin{eqnarray}
H_t &=& \frac{1}{\sqrt{q^2}}
\left\{(m_1^2-m_2^2)\, F_+ + q^2\, F_- \right\}\,,
\nonumber\\
H_\pm &=& 0\,,
\label{helS0b}\\
H_0 &=& \frac{2\,m_1\,|{\bf p_2}|}{\sqrt{q^2}} \,F_+ \,.
\nonumber
\end{eqnarray}

\vspace{0.5cm}
\noindent
(b) Spin $S=1$:

\begin{eqnarray}
H_t &=&
\frac{1}{m_1+m_2}\frac{m_1\,|{\bf p_2}|}{m_2\sqrt{q^2}}
\left\{ (m_1^2-m_2^2)\,(A_+ - A_0)+q^2 A_- \right\},
\nonumber\\
H_\pm &=&
\frac{1}{m_1+m_2}\left\{- (m_1^2-m_2^2)\, A_0
\pm 2\,m_1\,|{\bf p_2}|\, V \right\},
\label{helS1c}\\
H_0 &=&
\frac{1}{m_1+m_2}\frac{1}{2\,m_2\sqrt{q^2}}
\left\{-(m_1^2-m_2^2) \,(m_1^2-m_2^2-q^2)\, A_0
+4\,m_1^2\,|{\bf p_2}|^2\, A_+\right\}.
\nonumber
\end{eqnarray}
where $|{\bf p_2}|=\lambda^{1/2}(m_1^2,m_2^2,q^2)/(2\,m_1)$ 
is the momentum of the outgoing particles
in the $B_c$ rest frame.
%
\noindent
The semileptonic $B_c$-decay widths are given by

\begin{eqnarray*}
\Gamma(B_c^-\to M_{\bar cc}\, l \bar\nu) & = & \frac{G_F^2}{(2\,\pi)^3}
|V_{cb}|^2
\int\limits_{m_l^2}^{q^2_-} dq^2
\frac{(q^2-m_l^2)^2\,|{\mathbf p_2}|}{12\,m_1^2\,q^2}
\\
&&
\times\left\{
\left(1+\frac{m_l^2}{2\,q^2}\right)
\sum\limits_{i=\pm,0} \left(H_i^{B_c\to M_{\bar cc}}(q^2)\right)^2
+\frac{3\,m_l^2}{2\,q^2}\left(H_t^{B_c\to M_{\bar cc}}(q^2)\right)^2
\right\}\,,
\\
%
\Gamma(B_c^-\to \overline D^0\, l \bar\nu) & = & \frac{G_F^2}{(2\,\pi)^3}
|V_{ub}|^2
\int\limits_{m_l^2}^{q^2_-} dq^2
\frac{(q^2-m_l^2)^2\,|{\mathbf p_2}|}{12\,m_1^2\,q^2}
\\
&&
\times\left\{
\left(1+\frac{m_l^2}{2\,q^2}\right)
\sum\limits_{i=\pm,0} \left(H_i^{B_c\to \overline D^0}(q^2)\right)^2
+\frac{3\,m_l^2}{2\,q^2}\left(H_t^{B_c\to \overline D^0}(q^2)\right)^2
\right\}\,,
\\
\end{eqnarray*}
where $q^2=(m_1- m_2)^2$, $m_1\equiv m_{B_c}$, and
$m_2\equiv m_f$. Note that $M_{\bar cc}$ and $\overline D^0$ denote both
the pseudoscalar and vector cases.

\section{Numerical results}
\label{sec:results}
We take the following values~(\ref{eq:values}) of the meson masses
 and the $B_c$-meson's lifetime from the PDG \cite{Olive:2016xmw}.
 
\be
\def\arraystretch{1.5}
\begin{array}{ccccccc|c|c}
     m_{B_{c}}        &      m_{\eta_c}        &      m_{J/\psi}       	&     m_D 
	&m_{D^\ast}    	  &		 m_{D_s}		   &	  m_{D^{\ast}_{s}}  &					&	\tau_{B_{c}}  
\\\hline
 \ \ 6.275\ \   	  &  \ \ 2.984\ \   	   &  \ \ 3.097\ \  		&  \ \ 1.869\ \ 
 &\ \ 2.010\ \   	  &\ \ 1.968\ \  		   & \ \ 2.112\ \			& \ {\rm GeV} 		& \ \ 0.507 ps
\end{array}
\label{eq:values}
\en

The calculation of the semileptonic decay widths is straightforward. For the CKM-matrix elements we use
\begin{equation}
\def\arraystretch{2}
\begin{array}{ccccccc}
|V_{ud}|     & |V_{us}|      & |V_{cd}|    & |V_{cs}|     & |V_{cb}|       &
|V_{ub}| \\
\hline
\ \ 0.974 \ \ & \ \ 0.225 \ \  & \ \ 0.220\ \ & \ \ 0.995 \ \  & \ \ 0.0405 \ \ &
\ \ 0.00409 \\
\end{array}
\label{CKM}
\end{equation}

The value of the decay constant $f_{\eta_c}$ was calculated from the branching ratio for
the $\eta_c$ meson decay into two photons using the last data \cite{Olive:2016xmw}.
\noindent The quality of the fit may be assessed from the entries
in Table~\ref{tab:leptonic}.

\begin{table}[t]
\caption{Leptonic decay constants $f_H$ (MeV).}
\label{tab:leptonic}
\begin{center}
\def\arraystretch{1.5}
\begin{tabular}{|c|c|c|l|l|}
\hline
    & This work  & \hspace*{1cm} Other & \hspace*{1cm}Ref.  \\
\hline
$f_{B_c}$
&489  		& $489\pm4\pm3$ 		& LAT~\cite{Chiu:2007km}\\
&				& $395 \pm 15$ 			& \cite{Kiselev:2003uk} \\
\hline
$f_D$
&206 & $222.6\pm 16.7^{+2.8}_{-3.4}$  & CLEO \cite{Artuso:2005ym} \\
&      & $201 \pm 3 \pm 17$            & MILC LAT \cite{Aubin:2005ar} \\
&      & $235 \pm 8 \pm 14$            & LAT \cite{Chiu:2005ue} \\
&      & $210 \pm 10^{+17}_{-16}$      & UKQCD LAT \cite{Lellouch:2000tw}\\
&      & $211 \pm 14^{+2}_{-12}$       & LAT \cite{Becirevic:1998ua}\\
&      & 204.6$\pm$5.0                 & PDG~\cite{Agashe:2014kda}\\

$f_{D^\ast}$
&244  & $245 \pm 20^{+3}_{-2}$ 	& LAT \cite{Becirevic:1998ua}\\
&         & $278\pm13\pm10$ 			& LAT~\cite{Becirevic:2012ti}\\
&	 	   & $252.2\pm22.3\pm4$ 		& QCD SR~\cite{Lucha:2014xla}\\
\hline
$f_{D_s}$
 &257  & 257.5$\pm$4.6 		     & PDG~\cite{Agashe:2014kda}\\
 &     & $249 \pm 3 \pm 16$             & MILC LAT \cite{Aubin:2005ar} \\
 &     & $266 \pm 10\pm 18$             & LAT \cite{Chiu:2005ue}  \\
 &     & $290 \pm 20\pm 29\pm 29\pm 6$  & LAT \cite{Wingate:2003gm} \\
 &     & $236 \pm 8^{+17}_{-14}$        & UKQCD LAT \cite{Lellouch:2000tw}\\
 &     & $231 \pm 12^{+8}_{-1}$         & LAT \cite{Becirevic:1998ua}\\   
$f_{D^\ast_s}$
&272  	&311$\pm$9 						& LAT~\cite{Becirevic:2012ti}\\
& 			&$272(16)^{+3}_{-20}$ 			& LAT~\cite{Becirevic:1998ua}\\
&			&$305.5\pm26.8\pm5$ 			& QCD SR~\cite{Lucha:2014xla}\\
\hline
$\displaystyle\frac{f_{D_s}}{f_D}$
&1.25  		& 1.258$\pm$0.038 					& PDG~\cite{Agashe:2014kda}\\
&					& $1.24 \pm 0.01\pm 0.07$           & MILC LAT \cite{Aubin:2005ar} \\
&       			& $1.13 \pm 0.03\pm 0.05$         	& LAT \cite{Chiu:2005ue}  \\
&      			& $1.13 \pm 0.02^{+0.04}_{-0.02}$  	& UKQCD LAT \cite{Lellouch:2000tw}\\
&       			& $1.10 \pm 0.02$                 	& LAT \cite{Becirevic:1998ua}\\
\hline

$f_{\eta_c}$
&628  					&$420\pm 52$ 		& \cite{Hwang:1997ie} \\
&							&$337.7\pm 18.2$	& pQCD\cite{Sun:2017lla} \\
\hline
$f_{J/\psi}$
&415  					& $405\pm 14$		& pQCD\cite{Rui:2014tpa}\\
&							& $416.2\pm 7.4$	& pQCD\cite{Sun:2017lla} \\
\hline
\end{tabular}
\end{center}
\end{table}

\begin{figure}[ht]
\begin{center}
\hspace*{-0.5cm}
\begin{tabular}{lrr}
\includegraphics[width=0.30\textwidth]{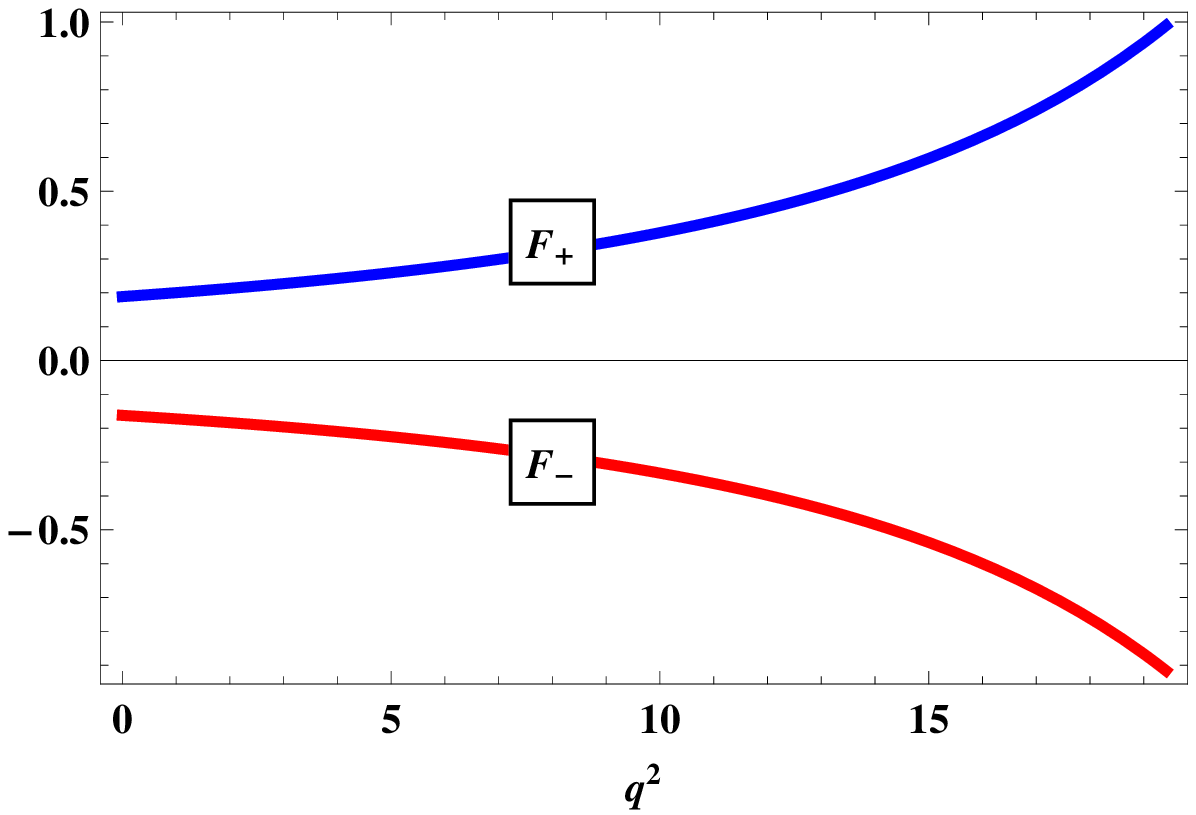} 
\includegraphics[width=0.30\textwidth]{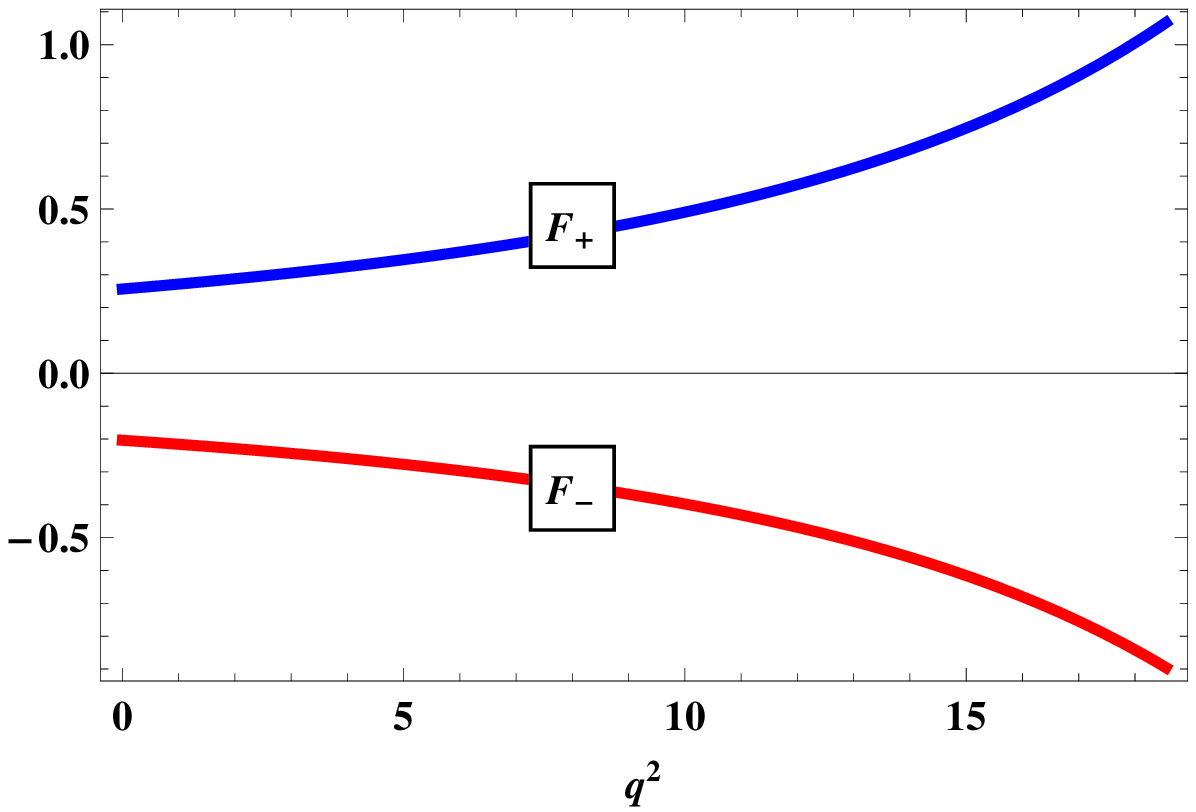} 
\includegraphics[width=0.30\textwidth]{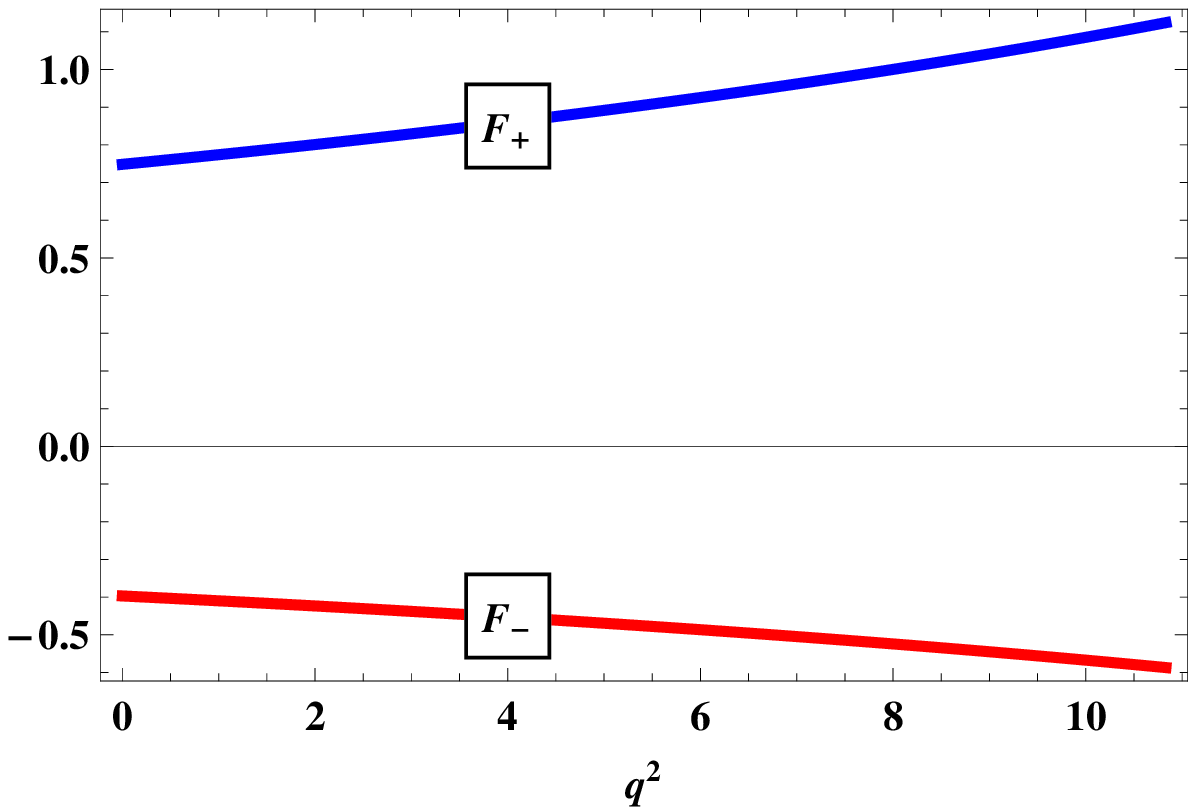} 
\end{tabular}
\end{center}
\caption{\label{fig:ffP}
The $F_+(q^2)$ and $F_-(q^2)$ form factors for $B_{c} \to D$,$B_{c} \to D_s$ and $B_{c} \to \eta_c$ transitions, respectively.  
}
\end{figure}

\begin{figure}[ht]
\begin{center}
\hspace*{-0.5cm}
\begin{tabular}{lrr}
\includegraphics[width=0.30\textwidth]{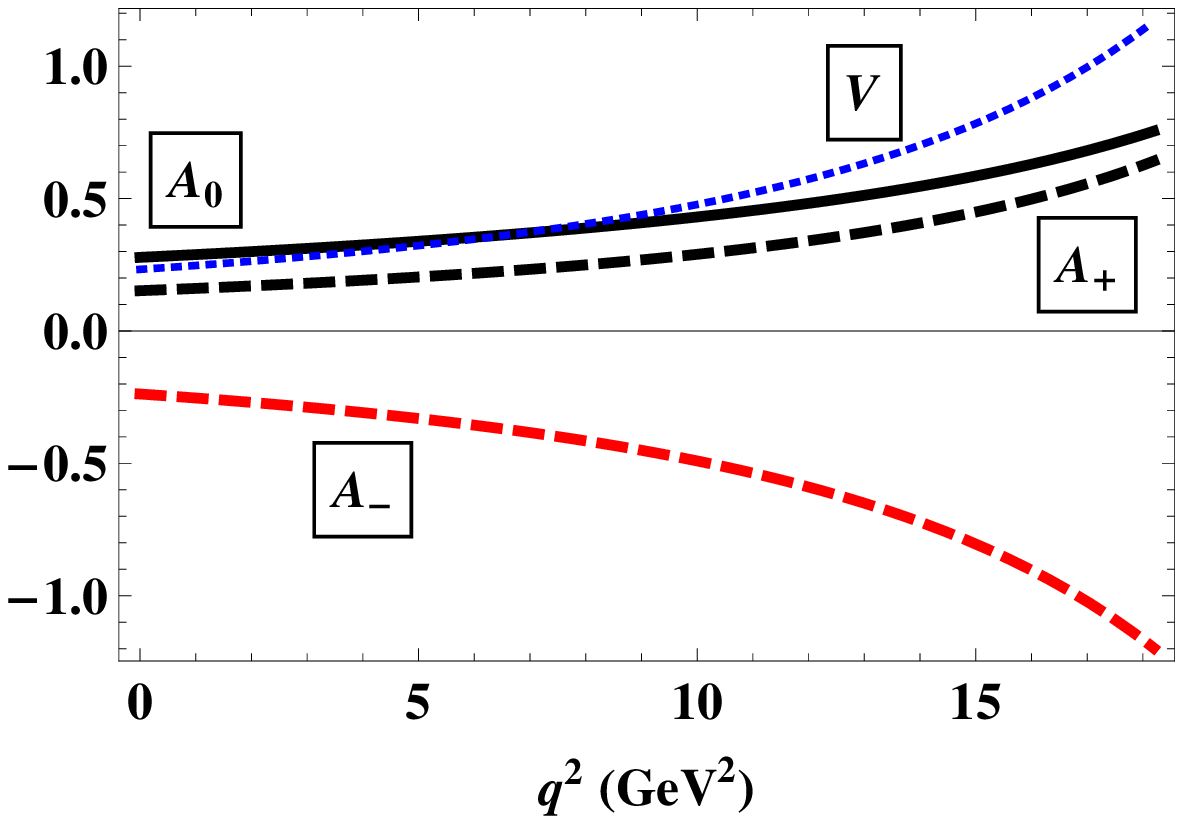} 
\includegraphics[width=0.30\textwidth]{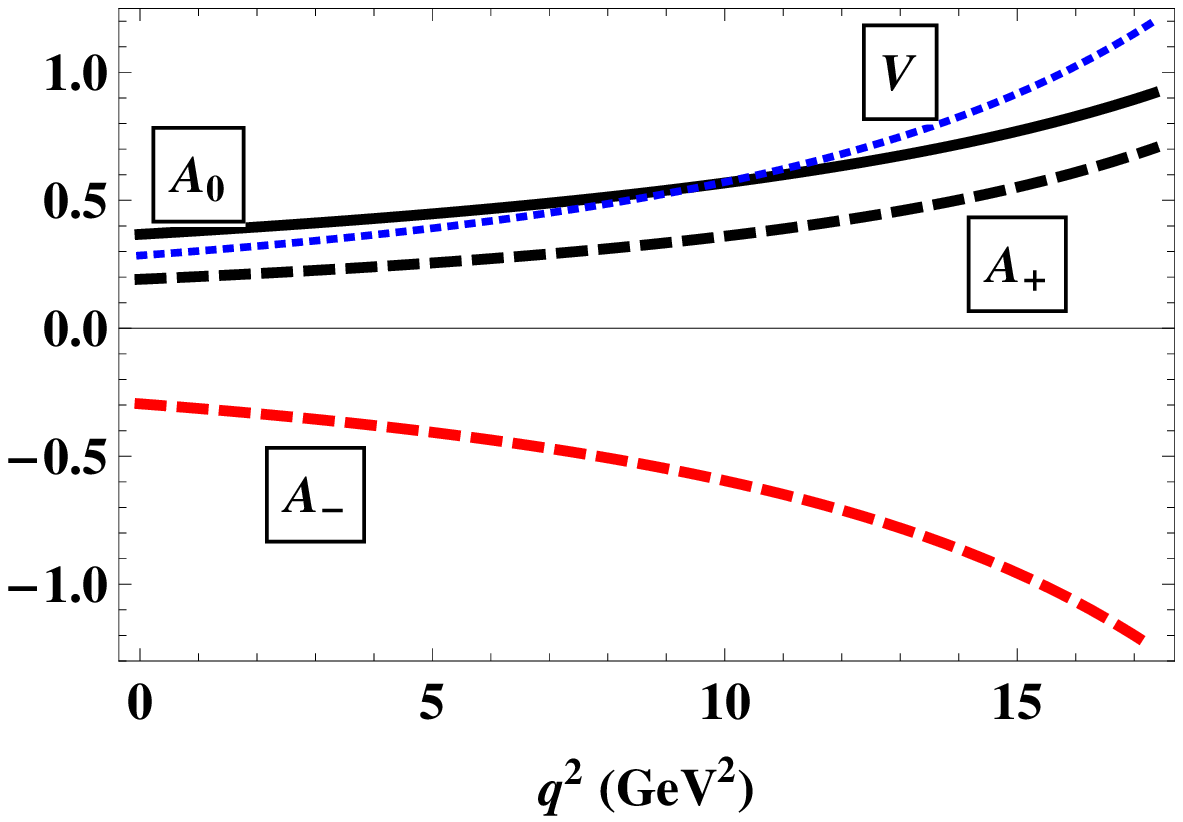} 
\includegraphics[width=0.30\textwidth]{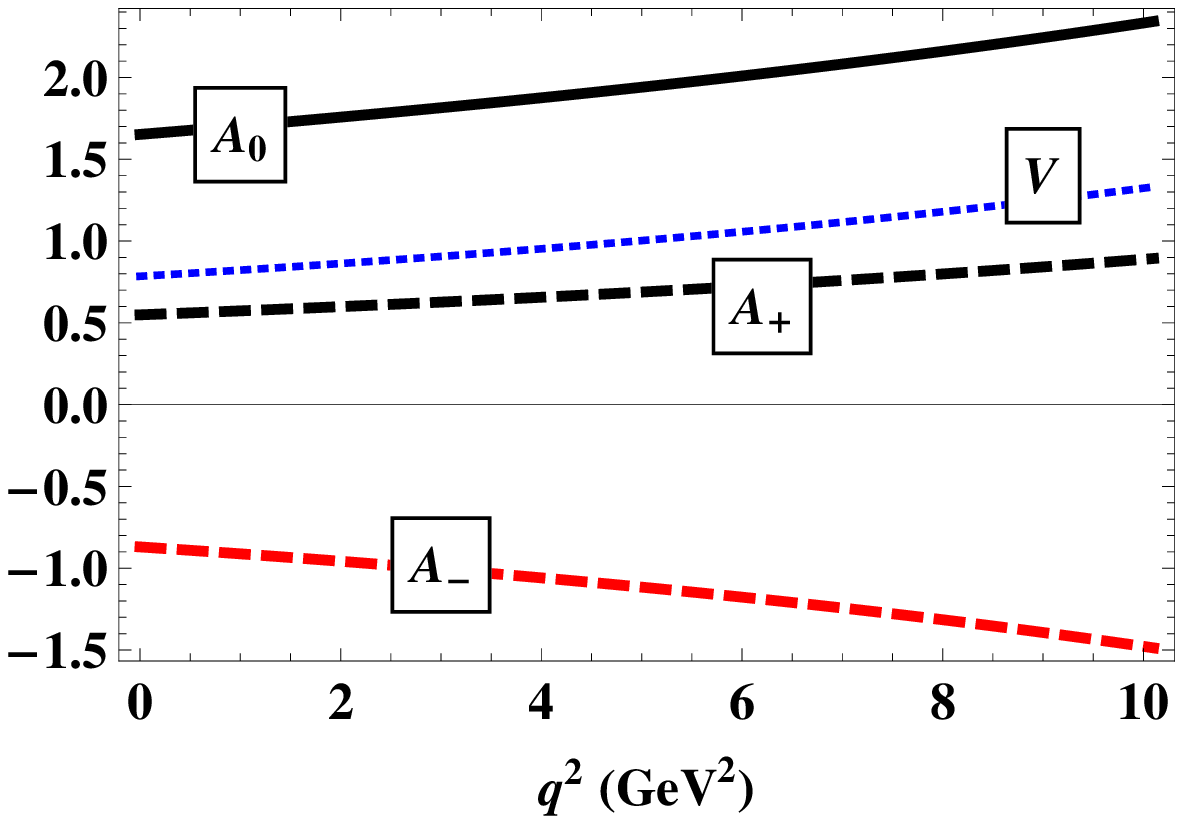} 
\end{tabular}
\end{center}
\caption{\label{fig:ffV}
The $A_0 , A_- , A_+ $ and $V$ form factors for $B_{c} \to D^{*}$, $B_{c} \to D_{s}^{*}$ and  $B_{c} \to J/\psi$ transitions, respectively.  
}
\end{figure}

The form factors are calculated in the full kinematical region of 
momentum transfer squared and are shown in Table~\ref{tab:ff}. The curves are depicted  in Fig.~\ref{fig:ffP} and Fig.~\ref{fig:ffV}.

\begin{table}[b]
\caption{
Form factors for $B_{c}\rightarrow D(D^{\ast})$, $B_{c}\rightarrow D_s(D_s^{\ast})$ and $B_{c}\rightarrow \eta_{c}(J/\psi)$ transitions.
Form factors are {\it approximated} by the form
$F(q^2)=F(0)/(1-a\,\hat s+b\,\hat s^2)$ with $\hat s=q^2/m_{B_c}^2$.
}
\vspace{0.3cm}
\begin{center}
\def\arraystretch{1.5}
\begin{tabular}{|c|c||c||c|}
\hline
   & { $B_{c}\rightarrow D(D^{\ast})     $} 
   & { $B_{c}\rightarrow D_s(D_s^{\ast}) $}
    & { $B_{c}\rightarrow \eta_{c}(J/\psi)$ } \\

\hline\hline
$F_{+}(0)$ &0.186  	&0.254    &0.74  \\ \hline
$F_{-}(0)$ &-0.160  &-0.202	  &-0.39\\ \hline\hline
$A_{0}(0)$ &0.276  	&0.365    &1.65\\ \hline
$A_{+}(0)$ &0.151  	&0.190    &0.55\\ \hline
$A_{-}(0)$ &-0.236  &-0.293   &-0.87\\ \hline
$V(0)$ &0.230  &0.282  &0.78\\ \hline
\end{tabular}
\label{tab:ff}
\end{center}
\end{table}

\vspace*{1cm}

\vspace*{1cm}
\clearpage
The results of our evaluation of the branching ratios of the 
semileptonic $B_c$ decays appear in
Table~\ref{tab:Bc-semlep}, which contains
our predictions for the  semileptonic $B_c$ decays into ground state
charmonium states and charm meson states. 
We compare our ratios of semileptonic decays of the $B_c$ meson with those of other models in Table~\ref{tab:ratio-semilep}.

\begin{table}[t]
\caption{\label{tab:Bc-semlep}
         Branching ratios (in $\%$) of
         semileptonic $B_c$ decays into ground state charmonium
         states.}
\def\arraystretch{1.5}
\begin{center}
\begin{tabular}{|l|l|l|l|l|l|l|l|}
\hline
 Mode & This work & \cite{Ivanov:2006ni} & \cite{Ivanov:2000aj} & \cite{KKL,exBc} & \cite{Chang:1992pt} &
 \cite{narod} &\cite{Wang:2012}\\
\hline

$B_c^- \to \eta_c \ell \nu$     &0.95 & 0.81 &  0.98  & 0.75 & 0.97 & 0.59   &0.44\\

$B_c^- \to \eta_c \tau \nu$  &0.24 & 0.22 & 0.27  & 0.23 &     & 0.20 &0.14\\
\hline

$B_c^- \to J/\psi \ell \nu $  & 1.67  & 2.07  &2.30  & 1.9  & 2.35 & 1.20  &1.01\\

$B_c^- \to J/\psi \tau \nu $ &0.40 & 0.49 &0.59   & 0.48 &     & 0.34 &0.29\\
\hline
 $B_c^- \to  \overline D^- \ell \nu $      &0.0033 &      0.0035    &0.018 & & 0.004 & 0.006 & 0.0032\\
 $B_c^- \to  \overline D^- \tau \nu $ &0.0021 &  0.0021 & 0.0094  & 0.002 &      &   &0.0022\\
\hline
 $B_c^- \to  \overline D^{\ast\,-} \ell \nu  $ &0.006 & 0.0038 & 0.034  && 0.018 & 0.018  & 0.011\\

 $B_c^- \to  \overline D^{\ast\,-} \tau \nu$ &0.0034  & 0.0022 & 0.019  & 0.008 &      &  &0.006\\
\hline
\end{tabular}
\end{center}
\end{table}

\begin{table}
\caption{\label{tab:ratio-semilep}
         Ratios of semileptonic decays of the $B_c$ meson}
\begin{center}
\begin{tabular}{|l|ccc|} 
\hline
	Decay rate 			& This work 		&\cite{Ivanov:2006ni}			&\cite{Wang:2012}	\\
	\hline
 $R_{\eta_c}=\frac{B_c^- \to \eta_c \ell \nu} {B_c^- \to \eta_c \tau \nu}$ 			&3.96	&3.68	&3.2	\\
 \hline
 $R_{J/\psi}=\frac{B_c^- \to J/\psi \ell \nu} {B_c^- \to J/\psi \tau \nu}$  		&4.18	&4.22	&3.4	\\
 \hline
 $R_{ D}=\frac{B_c^- \to D^- \ell \nu} {B_c^- \to D^- \tau \nu}$ 					&1.57	&1.67	&1.42	\\
 \hline
 $R_{ D^{\ast}}=\frac{B_c^- \to D^{\ast\,-} \ell \nu} {B_c^- \to D^{\ast\,-} \tau \nu}$ 	&1.76	&1.72	&1.66	\\
\hline
\end{tabular}
\end{center}
\end{table}

\section*{Acknowledgment}

Authors A. Issadykov, M.A. Ivanov and G.S. Nurbakova acknowledge the partial support by the
Ministry of Education and Science of the Republic of Kazakhstan, grant 3092/GF4, state registration No. 0115RK01040. 

Author A. Issadykov is grateful for the support by the JINR, grant number 17-302-03.


\begin{thebibliography}{99}

\bibitem{Efimov:1988yd}
  G.~V.~Efimov and M.~A.~Ivanov,
  Int.\ J.\ Mod.\ Phys.\ A {\bf 4} (1989) 2031.
  doi:10.1142/S0217751X89000832
   
\bibitem{Faessler:2002ut}
  A.~Faessler, T.~Gutsche, M.~A.~Ivanov, J.~G.~Korner and V.~E.~Lyubovitskij,
  Eur.\ Phys.\ J.\ direct {\bf 4} (2002) no.1,  18
  doi:10.1007/s1010502c0018
  [hep-ph/0205287].
  
\bibitem{Branz:2009cd}
  T.~Branz, A.~Faessler, T.~Gutsche, M.~A.~Ivanov, J.~G.~Korner and V.~E.~Lyubovitskij,
  Phys.\ Rev.\ D {\bf 81} (2010) 034010
  doi:10.1103/PhysRevD.81.034010
  [arXiv:0912.3710 [hep-ph]].

\bibitem{Weinberg:1962hj}
  S.~Weinberg,
  Phys.\ Rev.\  {\bf 130} (1963) 776.
  doi:10.1103/PhysRev.130.776
  
\bibitem{Salam:1962ap}
  A.~Salam,
  Nuovo Cim.\  {\bf 25} (1962) 224.
  doi:10.1007/BF02733330
   
\bibitem{Dubnicka:2016nyy}
  S.~Dubni\v{c}ka,A.Z.~Dubni\v{c}kov\'{a}, A.~Issadykov, M.~A.~Ivanov, A.~Liptaj and S.~K.~Sakhiyev,
  Phys.\ Rev.\ D {\bf 93} (2016) no.9,  094022
  doi:10.1103/PhysRevD.93.094022
  [arXiv:1602.07864 [hep-ph]].

\bibitem{Ivanov:2000aj}
  M.~A.~Ivanov, J.~G.~Korner and P.~Santorelli,
  Phys.\ Rev.\ D {\bf 63} (2001) 074010
  doi:10.1103/PhysRevD.63.074010
  [hep-ph/0007169].

\bibitem{Olive:2016xmw}
  C.~Patrignani {\it et al.} [Particle Data Group],
  Chin.\ Phys.\ C {\bf 40} (2016) no.10,  100001.
  doi:10.1088/1674-1137/40/10/100001

\bibitem{Chiu:2007km}
  T.~W.~Chiu {\it et al.} [TWQCD Collaboration],
  Phys.\ Lett.\ B {\bf 651} (2007) 171
  doi:10.1016/j.physletb.2007.06.017
  [arXiv:0705.2797 [hep-lat]].
      
\bibitem{Kiselev:2003uk}
  V.~V.~Kiselev,
  Central Eur.\ J.\ Phys.\  {\bf 2}, 523 (2004)
  [arXiv:hep-ph/0304017].

\bibitem{Artuso:2005ym}
  M.~Artuso {\it et al.}  [CLEO Collaboration],
  Phys.\ Rev.\ Lett.\  {\bf 95} (2005) 251801
  [arXiv:hep-ex/0508057].

\bibitem{Aubin:2005ar}
  C.~Aubin {\it et al.},
  Phys.\ Rev.\ Lett.\  {\bf 95} (2005) 122002
  [arXiv:hep-lat/0506030].

\bibitem{Chiu:2005ue}
  T.~W.~Chiu, T.~H.~Hsieh, J.~Y.~Lee, P.~H.~Liu and H.~J.~Chang,
  Phys.\ Lett.\ B {\bf 624}, 31 (2005)
  [arXiv:hep-ph/0506266].
  
\bibitem{Lellouch:2000tw}
  L.~Lellouch and C.~J.~D.~Lin  [UKQCD Collaboration],
  Phys.\ Rev.\ D {\bf 64}, 094501 (2001)
  [arXiv:hep-ph/0011086].
  
\bibitem{Becirevic:1998ua}
  D.~Becirevic, P.~Boucaud, J.~P.~Leroy, V.~Lubicz, G.~Martinelli, F.~Mescia and F.~Rapuano,
  Phys.\ Rev.\ D {\bf 60}, 074501 (1999)
  [arXiv:hep-lat/9811003].
   
\bibitem{Agashe:2014kda}
  K.~A.~Olive {\it et al.} [Particle Data Group],
  Chin.\ Phys.\ C {\bf 38} (2014) 090001.
  doi:10.1088/1674-1137/38/9/090001

\bibitem{Becirevic:2012ti}
  D.~Becirevic, V.~Lubicz, F.~Sanfilippo, S.~Simula and C.~Tarantino,
  JHEP {\bf 1202} (2012) 042
  doi:10.1007/JHEP02(2012)042
  [arXiv:1201.4039 [hep-lat]].
  
\bibitem{Lucha:2014xla}
  W.~Lucha, D.~Melikhov and S.~Simula,
  Phys.\ Lett.\ B {\bf 735} (2014) 12
  doi:10.1016/j.physletb.2014.06.007
  [arXiv:1404.0293 [hep-ph]].

\bibitem{Wingate:2003gm}
  M.~Wingate, C.~T.~H.~Davies, A.~Gray, G.~P.~Lepage and J.~Shigemitsu,
  Phys.\ Rev.\ Lett.\  {\bf 92}, 162001 (2004)
  [arXiv:hep-ph/0311130].


\bibitem{Hwang:1997ie}
  D.~S.~Hwang and G.~H.~Kim,
  Z.\ Phys.\ C {\bf 76}, 107 (1997)
  [arXiv:hep-ph/9703364].
  
\bibitem{Sun:2017lla}
  J.~Sun, Y.~Yang, N.~Wang, J.~Huang and Q.~Chang,
  Phys.\ Rev.\ D {\bf 95} (2017) no.3,  036024
  doi:10.1103/PhysRevD.95.036024
  [arXiv:1703.00155 [hep-ph]].

\bibitem{Rui:2014tpa}
  Z.~Rui and Z.~T.~Zou,
  Phys.\ Rev.\ D {\bf 90} (2014) no.11,  114030
  doi:10.1103/PhysRevD.90.114030
  [arXiv:1407.5550 [hep-ph]].

\bibitem{Ivanov:2006ni}
  M.~A.~Ivanov, J.~G.~Korner and P.~Santorelli,
  Phys.\ Rev.\ D {\bf 73} (2006) 054024
  doi:10.1103/PhysRevD.73.054024
  [hep-ph/0602050].

\bibitem{KKL}
V.~V.~Kiselev, A.~E.~Kovalsky and A.~K.~Likhoded,
Nucl.\ Phys.\ B {\bf 585}, 353 (2000) [arXiv:hep-ph/0002127];
arXiv:hep-ph/0006104.

\bibitem{exBc}
V.~V.~Kiselev,
arXiv:hep-ph/0211021.

\bibitem{Chang:1992pt}
C.~H.~Chang and Y.~Q.~Chen,
Phys.\ Rev.\ D {\bf 49}, 3399 (1994).


\bibitem{narod}
A.~Y.~Anisimov, I.~M.~Narodetsky, C.~Semay and B.~Silvestre-Brac,
Phys.\ Lett.\ B {\bf 452}, 129 (1999) [arXiv:hep-ph/9812514];
A.~Y.~Anisimov, P.~Y.~Kulikov, I.~M.~Narodetsky and
K.~A.~Ter-Martirosian,
Phys.\ Atom.\ Nucl.\  {\bf 62}, 1739 (1999) [Yad.\ Fiz.\  {\bf
62}, 1868 (1999)] [arXiv:hep-ph/9809249].

\bibitem{Wang:2012} 
  W.~F.~Wang, Y.~Y.~Fan and Z.~J.~Xiao,
  Chin.\ Phys.\ C {\bf 37} (2013) 093102
  doi:10.1088/1674-1137/37/9/093102
  [arXiv:1212.5903 [hep-ph]].
  


\end{thebibliography}
\end{document}